\documentclass{article}

\usepackage{graphicx} 
\usepackage{amsmath}
\usepackage{enumitem}
\usepackage{amsfonts}
\usepackage{tikz}
\usepackage{hyperref}
\usepackage{float}
\usepackage{array}
\usepackage{url}
\usepackage{breakurl}
\usepackage[margin=1in, textwidth=6.5in]{geometry} 
\usepackage{xurl}
\usepackage{xcolor}
\usepackage[normalem]{ulem}
\usepackage{cleveref}
\crefname{figure}{Figure}{Figures}
\crefrangeformat{figure}{Figures~#3#1--#2#4}

\title{The Effect of Assortativity on Mpox Spreading with Two Core Groups}
\author{Fanni Ned\'enyi$^{1}$\thanks{\texttt{fanni.nedenyi@hcemm.eu}}, J\'anos M. Benke$^{2,3}$\thanks{\texttt{jbenke@math.u-szeged.hu}}, and Gergely R\"ost$^{1,2,3}$\thanks{\texttt{rost@math.u-szeged.hu}}}

\date{\today}

\begin{document}
\maketitle

\begin{center}
$^{1}$ Scientific Computing Advanced Core Facility, Hungarian Center of Excellence for Molecular Medicine (HCEMM), Szeged, Hungary\\
$^{2}$ Bolyai Institute, University of Szeged, 6720 Szeged, Hungary\\
$^{3}$ National Laboratory for Health Security, University of Szeged, 6720 Szeged, Hungary\\
\end{center}

\begin{abstract}
The spread of infectious diseases often concentrates within specific subgroups of a broader population. For instance, during recent mpox outbreaks in non-endemic countries, transmission primarily affected men who have sex with men (MSM). However, the internal structure of these subpopulations plays a crucial role in disease dynamics and should be accurately represented in mathematical models. In this study, we highlight the importance of modeling interactions between distinct subgroups and their impact on transmission patterns. We consider a stochastic SEIR-based model with two core groups embedded into the general population, and investigate the outcome of the outbreak with different levels of symmetry between these groups and assortativity in their contacts.
Our results indicate that the efficiency of commonly used non-pharmaceutical interventions is greatly influenced by these factors, hence they should be considered in the design of intervention strategies.
\end{abstract}

\section{Introduction}
\label{sec:1}
Mpox is a viral zoonosis with symptoms similar to those of smallpox, but causes a less severe illness with much lower mortality. The virus, a double-stranded DNA virus from the Poxviridae family, was first identified in 1958 in a laboratory monkey at the Statens Serum Institute in Copenhagen, Denmark. The first human case was reported in a young child in the Democratic Republic of the Congo (DRC) in 1970 \cite{WHO1}. Mpox is typically transmitted to humans through contact with biological fluids or lesions of an infected animal, often a rodent \cite{Gessain}. Human-to-human transmission occurs via direct contact with infected biological fluids, skin lesions, mucous membranes, or through inhalation of contaminated respiratory droplets \cite{Beeson}. The incubation period generally ranges from 6 to 13 days but can extend from 5 to 21 days \cite{WHO1, WHO2}.

For decades, mpox was primarily a local concern in Africa, where transmission was driven by environmental factors such as poor living conditions and frequent contact with infected rodents. However, the global outbreak in May 2022 saw the virus spread to numerous non-endemic countries, with cases disproportionately affecting a close-knit subpopulation—primarily men who have sex with men (MSM). This shift in transmission dynamics highlights the importance of understanding how the structure of such subpopulations influences disease spread \cite{pan}.

A variety of mathematical models have been developed to analyze mpox transmission dynamics and evaluate the effectiveness of various intervention strategies. These models have been used to assess the impact of vaccination, isolation, and contact tracing in both high- and low-risk groups, as well as the role of mass gatherings, public awareness campaigns, and other control measures \cite{yuan, mansouri, uchenna}. Given the disproportionate burden of mpox among certain subpopulations, several studies have placed special emphasis on modeling core groups, which play a central role in sustaining transmission. For instance, studies such as \cite{savinkina, omame} have specifically examined how targeted interventions within core groups—such as men who have sex with men (MSM)—can influence overall outbreak dynamics. These models highlight the importance of accounting for heterogeneity in contact patterns and behavioral factors when designing effective public health responses.

In this study, we investigate how the internal structure of two interconnected core groups within the general population influences mpox transmission dynamics. Specifically, we assess the effectiveness of commonly implemented non-pharmaceutical interventions (NPIs), such as contact reduction, in mitigating disease spread. While our previous research \cite{mpox_sajat} modeled the MSM subpopulation as a single, homogeneous group, we refine this approach by introducing a more detailed structure in which the subpopulation is divided into two distinct groups with different interaction patterns. This distinction allows us to capture more realistic transmission pathways and assess how the nature of interactions between these groups impacts overall outbreak progression. By incorporating this additional complexity, our study provides deeper insights into the role of subgroup heterogeneity in disease spread and intervention effectiveness.

Although our analysis focuses on mpox, the insights gained from this study are applicable to any infectious disease with a host population with core groups structured similarly to the MSM subpopulation in mpox outbreaks. However, for the purposes of this research, we calibrate our model parameters specifically to mpox to ensure our findings more closely reflect real-world disease behavior.

\section{Population dynamics}
\subsection{SEIR Model Structure}
Consider a population of constant size $N$. Our stating point is the SEIR approach, meaning that at any time we keep track of the number of individuals in each disease state, which are:
\begin{itemize}
\item $S$, the number of susceptible individuals (those able to contract the disease);
\item $E$, the number of exposed individuals (infected but  not yet infectious);
\item $I$, the number of infectious individuals (those capable of transmitting the disease);
\item $R$, the number of recovered individuals (assumed to become fully immune).
\end{itemize}
Each individual, at any time, is in exactly one of those compartments. Thus, $N=S(t)+E(t)+I(t)+R(t)$ is unchanged over time.

\begin{figure}
    \centering
    \includegraphics[width=\textwidth]{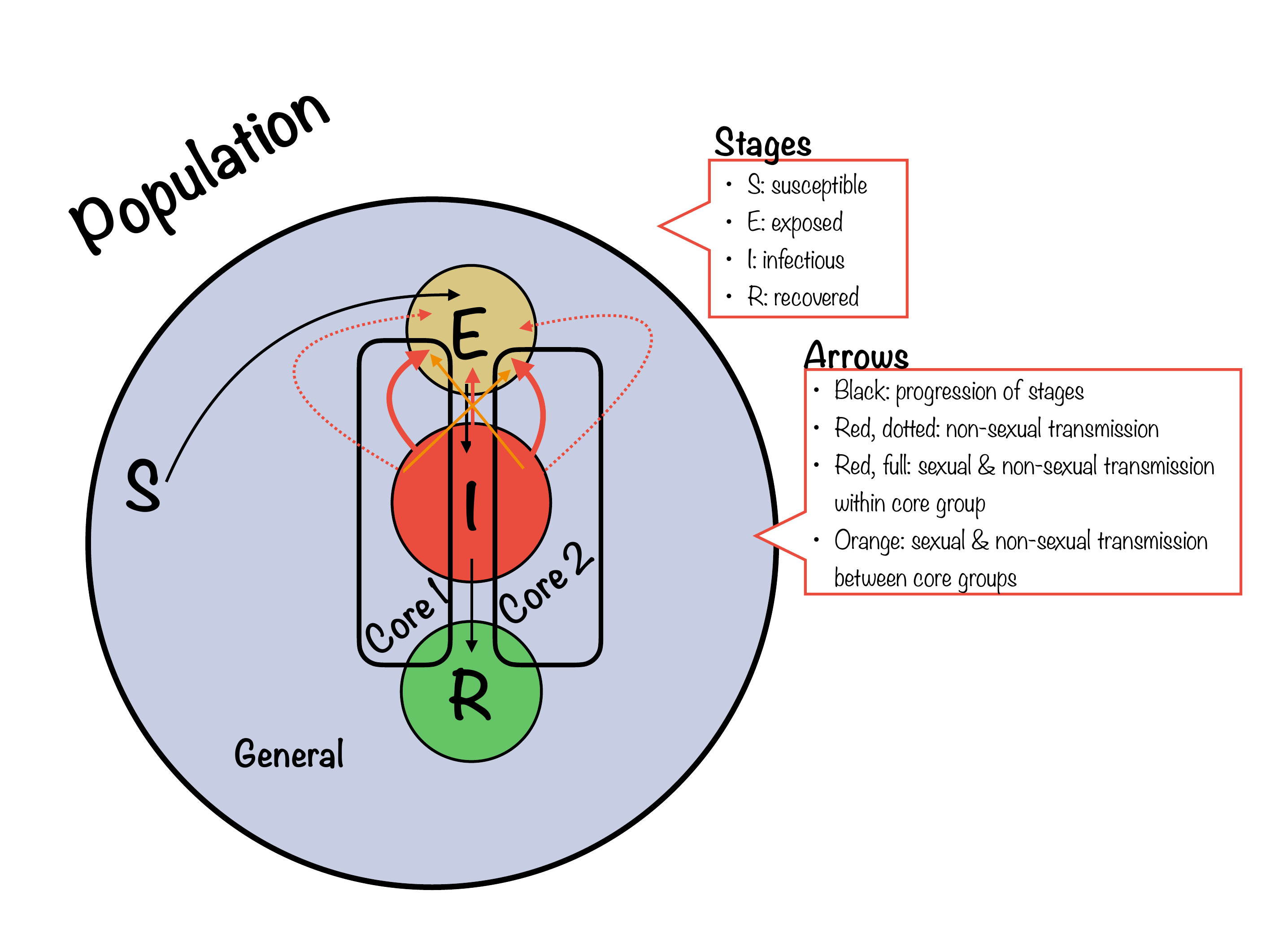}
    \caption{Illustation of our model structure, with two core groups (Core 1 and Core 2) embedded into the general population. Black arrows represent transitions between compartments, red and orange arrows represent transmission pathways.}
    \label{fig:dynamics}
\end{figure}

Figure~\ref{fig:dynamics} shows the structure of the population.
All of the individuals may contact each other, but the number of their contacts and the ways a potential transmission can happen differ. In particular,
\begin{itemize}
\item susceptible individuals can be exposed by an infectious host;
\item recovered individuals can no longer spread or contract the disease;
\item exposed individuals become infectious after an exponentially distributed amount of time (with parameter $\delta$);
\item infectious individuals
\begin{itemize}
\item recover after an exponentially distributed  amount of time (having parameter $\gamma$),
\item during their infectious period they infect susceptible individuals with the virus with rate $\beta$. 
\end{itemize}
\end{itemize}

This latter transmission parameter $\beta$ has a crucial role in determining the reproduction number of the virus. However, depending on which group and individual belonging to, they infect others from different groups with different likelihood. The next section explains this parameter in depth.

\subsection{Interactions Among Groups}\label{sec:dynamics}

This subsection is about the structure and relationship among the general population and the two core groups, which has been tailored to fit the real-life spread of mpox disease. As it has already been explained in the Introduction, in
non-endemic countries mpox tended to spread mainly within the high-risk MSM subpopulation. This is due
to the fact that this particular illness mostly spreads through sexual contacts and only secondarily via other
types of contacts including close contact, respiratory droplets, aerosols, and fomites. Moreover, the MSM
community is known to have a closely interconnected social and sexual network, and this higher density
of contacts facilitates the virus to spread more readily. In our model we take into consideration two types of
transmissions: sexual and non-sexual. The following assumptions establish our mathematical model:
\begin{itemize}
\item The number of non-sexual contacts does not depend on the type of the individual.
\item The non-sexual contacts follow random mixing. For any individual, the proportion of general individuals among their contacts is the same as in the entire population. Thus if a transmission happens through non-sexual contact, then the probability that the newly exposed individual is a general one is the same as the proportion of general ones in the population.
\item An individual only has sexual contact with an individual of similar type (general with general, core with core).
\item For an individual in one of the core groups its sexual contacts are divided between its own core group and the other one. The proportion of sexual contacts within its own group is denoted by $cg_{own}^{(1)}$ and $cg_{own}^{(2)}$ for Core Group 1 and Core Group 2, respectively.
\end{itemize}
These assumptions mean that non-sexual contacts are completely random, while sexual contacts are in-group. In the case of core groups, we measure what fraction of those are in very same core group. These considerations lead us to the following $\beta$ values. Instead of a homogeneous $\beta$ value, we break down the expected number of daily infections by the group of the infectious (lower index) and that of the potentially exposed (upper index) individuals. We introduce the following notations:
\begin{itemize}
    \item $p^{\textrm{s}}$ and $p^{\textrm{ns}}$ denote the probabilities of the sexual and non-sexual transmissions, respectively. This means that if a sexual or non-sexual contact occurs between two individuals, where one is able to infect the other, then this infection happens with probability $p^{\textrm{s}}$ and $p^{\textrm{ns}}$, respectively.
    \item $c^{\textrm{ns}}$ denotes the number of expected non-sexual contacts per day.
    \item $c^{\textrm{s}}_{\textrm{gp}}$ and $c^{\textrm{s}}_{\textrm{cg}}$ denote the number of expected sexual contacts in the general population and in the core groups, respectively, per day.
    \item  $q_{\mathrm{gp}}$, $q_{\mathrm{cg^{1}}}$ and  $q_{\mathrm{cg^{2}}}$ denote the proportion of the general and the core groups within the entire population, respectively (hence $q_{\mathrm{gp}}+q_{\mathrm{cg^{1}}}+q_{\mathrm{cg^{2}}}=1$).
\end{itemize}
Therefore, from the $\beta_a^b$ parameters, which, again, denote the expected number of people from group $b$  infected by an infectious group $a$ individual per day, we can build the following who-infects-who matrix (on a per day basis):

\begin{equation}\label{betas}
\begin{split}
&
\mathcal{B} = \left[
\begin{array}{ccc}
    \beta_{\mathrm{gp}}^{\mathrm{gp}} & \beta_{\mathrm{gp}}^{\mathrm{cg^{1}}} & \beta_{\mathrm{gp}}^{\mathrm{cg^{2}}} \\ 
    \beta_{\mathrm{cg^{1}}}^{\mathrm{gp}} & \beta_{\mathrm{cg^{1}}}^{\mathrm{cg^{1}}} & \beta_{\mathrm{cg^{1}}}^{\mathrm{cg^{2}}} \\ 
    \beta_{\mathrm{cg^{2}}}^{\mathrm{gp}} & \beta_{\mathrm{cg^{2}}}^{\mathrm{cg^{1}}} & \beta_{\mathrm{cg^{2}}}^{\mathrm{cg^{2}}}
\end{array}
\right] 
\\
&
=
\left[
\begin{array}{ccc}
    c^{\textrm{s}}_{\textrm{gp}} p^{\textrm{s}} + c^{\textrm{ns}} p^{\textrm{ns}} q_{\textrm{gp}} & c^{\textrm{ns}}               p^{\textrm{ns}} q_{\textrm{cg}^{1}} & c^{\textrm{ns}}               p^{\textrm{ns}} q_{\textrm{cg}^{2}} \\ 
    c^{\textrm{ns}}               p^{\textrm{ns}} q_{\textrm{gp}} & c^{\textrm{s}}_{\textrm{cg}} p^{\textrm{s}} cg_{own}^{(1)} +         c^{\textrm{ns}} p^{\textrm{ns}} q_{\textrm{cg}^{1}} & c^{\textrm{s}}_{\textrm{cg}} p^{\textrm{s}} (1 - cg_{own}^{(1)})     + c^{\textrm{ns}} p^{\textrm{ns}} q_{\textrm{cg}^{2}} \\ 
    c^{\textrm{ns}}               p^{\textrm{ns}} q_{\textrm{gp}} & c^{\textrm{s}}_{\textrm{cg}} p^{\textrm{s}} (1 - cg_{own}^{(2)})       + c^{\textrm{ns}} p^{\textrm{ns}} q_{\textrm{cg}^{1}} & c^{\textrm{s}}_{\textrm{cg}} p^{\textrm{s}} cg_{own}^{(2)} + c^{\textrm{ns}} p^{\textrm{ns}} q_{\textrm{cg}^{2}}
\end{array}
\right].
\end{split}
\end{equation}

Here we also introduce the notations for the reproduction numbers of the individuals. The basic reproduction number is the expected number of secondary infections generated by an infectious individual during its infectious phase in a susceptible population, and is usually denoted by $R_0$. However, due to the heterogeneity in sexual contacts, we define reproduction numbers for various types. We are particularly interested in the expected number of people infected by an infectious individual during its infectious phase within its own group. To this end, we introduce $R_0^{\textrm{CG1}}$, $R_0^{\textrm{CG2}}$ and $R_0^{\textrm{GP}}$, representing the reproduction numbers for individuals in the core groups (CG1 and CG2) and the general population (GP) within their own group, respectively. These notations help us differentiate and track the reproductive rates within each group separately, which is crucial for understanding the dynamics of disease spread in structured populations.

Since the expected duration of infectious period is $\gamma^{-1}$ for all infected individuals, we can construct a next generation matrix as $\mathcal{N}=\mathcal{B}\cdot \rm{diag} ({\gamma^{-1},\gamma^{-1},\gamma^{-1}})$.

\subsection{The development of the population}
We have defined all components in the previous subsections to run a stochastic simulation of an outbreak. Typically, we start with a small number of infected individuals, who infect others in different groups based on the matrix $\mathcal{B}$ until they recover. Newly infected individuals spend some time in the exposed class, then move to the infected compartment and start to infect others. When calculating new infections, we have take into account if the contacted individual is susceptible or not.

Next, we describe the temporal development of the population, specifically the simulation process. We use the Gillespie algorithm \cite{Gillespie1976}, which is a widely-used stochastic simulation method for modeling biochemical systems and population dynamics. The algorithm allows for the simulation of the system's state changes over time by considering the probabilistic nature of events.

Each time period leading to an action (such as becoming infectious, infecting a new individual, or recovering) follows an exponential distribution with a specific parameter. Consequently, the time before the next event in the population (whether an exposure, someone becoming infectious, or recovering) is determined by the minimum of several exponential random variables, which itself is exponentially distributed with a parameter equal to the sum of the individual rates. We then determine the type of action by further breaking down this aggregated parameter. Finally, we pinpoint the exact action by refining this breakdown even further. For a more detailed description of the process see the Appendix of the paper \cite{mpox_sajat}.

A simulation process is terminated in the event of extinction, which occurs when the epidemic ends, meaning there are no actively exposed or infectious individuals left in the entire population.

\subsection{Parameters}
\label{sec:par}
For the purposes of this paper we, will focus on two special cases of the above-described general setup.

\subsubsection{Symmetric core groups}

The first case represents a setup where the two groups have a symmetric role, and we assume that the two core groups have the same size, i.e. $q_{\mathrm{cg^{1}}}=q_{\mathrm{cg^{2}}}$. Since sexual contacts are symmetric, in terms of the $cg_{own}$ parameters, 
$$cg_{own}^{(1)} = cg_{own}^{(2)}$$
holds to keep the balance between the contacts of the two groups.
For example, if $cg_{own}^{(1)} = cg_{own}^{(2)} = 0.8$, then $80\%$ of its sexual connections are within its own core group, the rest are with the other core group. We can vary this parameter, but the role of the two groups remain symmetric. Assortative mixing occurs when individuals are more likely to interact with others who have similar characteristics. In this case, when $cg_{own}^{(1)}$ is large, individuals are mostly establish sexual encounters with others from the same core group, rather than from the other core group. Conversely, for small values of $cg_{own}^{(1)}$, the mixing with respect to the core groups becomes disassortative. In any case, their roles remain symmetric.

\subsubsection{Asymmetric core groups}

For comparison, we consider the asymmetric case too. For example, if  $cg_{own}^{(1)} = 1 -cg_{own}^{(2)} = 0.8$, then, for any core individual, no matter which core group it is in, $80\%$ of its sexual connections are within the first core group, the rest are with the second core group. In order to fit this asymmetrical case into our model, we need to be mindful of the sizes of the core groups. Indeed, as we wish to have the same number of sexual contacts for each core individual, we have to make sure that the asymmetry of the relationship between the two groups does not harm this assumption.

To determine the relative sizes of the two core groups, the daily sexual contact number ($c^\mathrm{s}$) has to be equal to the number of sexual contacts attributed to a Core Group 1 individual, and we have the balance law
\begin{equation}
c^\mathrm{s} = \frac{N_1 c^\mathrm{s} cg_{own}^{(1)} + N_2 c^\mathrm{s} (1-cg_{own}^{(2)})}{N_1},
\end{equation}
which results in $N_2= N_1/4$ for $cg_{own}^{(1)} = 0.8$. 

\begin{table}
\centering
\begin{tabular}{p{2cm} p{2cm} p{5cm} p{2cm}}
 & Notation & Definition & Value \\
\hline
Demographic
 & $N$ & Total size of the population & $10 \; 000$ \\
 related & $q_{\mathrm{gp}}$ & Proportion of the general population & 0.95 \\
 & $q_{\mathrm{cg^{(1)}}}$ & Proportion of core group 1& Dynamic \\
 & $q_{\mathrm{cg^{(2)}}}$ & Proportion of core group 2& Dynamic \\
 & $c^\mathrm{ns}$& Number of non-sexual contacts (day$^{-1}$) & 10 \\
 & $c^\mathrm{s}_\mathrm{gp}$ & Number of sexual contacts in the GP (day$^{-1}$) & 0.125 \\
 & $c^\mathrm{s}_\mathrm{cg}$ & Number of sexual contacts in the CGs (day$^{-1}$) & 1 \\
\hline
Transmission 
 & $p^\mathrm{ns}$& Probability of non-sexual transmission& 0.00125 \\
 related & $p^\mathrm{s}$ & Probability of sexual transmission& 0.1 \\
 & $\delta$ & Rate parameter of the exponentially distributed latent period (day$^{-1}$) & 0.1 \\
 & $\gamma$ & Rate parameter of the exponentially distributed infectious period (day$^{-1}$) & 0.05 \\
\hline
Simulation
 & $m$ & Number of simulation runs & $1 \;000$ \\
 related & $i$ & Number of initial infectious individuals in the CG 1 & $10$ \\
\end{tabular}
\caption{Parameter values.}
\label{parTable}
\end{table}

Therefore, for the asymmetric case with $cg_{own} = 0.8$, Core Group 1 has to be 4 times the size of Core Group 2, meaning that, since together they have 500 individuals, Core Group 1 is of size 400, while Core Group 2 is of size 100.

The values of the parameters we used are summarized in Table \ref{parTable}. The parameters $p^\mathrm{s}$, $\delta$, and $\gamma$ for mpox were suggested by previous articles \cite{Bra, WHO4}.
The number of initial infectious individuals in the Core Group 1 $i$ is equal to 10, because in this case the approximated probability that the process dies out immediately in the core group (approximately $\left(1/R_0^\textrm{CG1}\right)^i$) is negligible.
The rest of the parameters were chosen to produce epidemiologically meaningful and feasible results.

Next, in \eqref{R0s}, we present the reproduction numbers based on the above defined specific parameters depending on the variables $cg_{own}^{(1)}$ and $cg_{own}^{(2)}$, which will have multiple values for our simulation runs. We show a next generation matrix of $3 \times 3$ similarly to when we presented the $\beta$ values in \eqref{betas}. Note that the entries are the corresponding values of \eqref{betas} divided by $\gamma$, to obtain reproduction numbers, hence they are

\begin{equation}\label{R0s}
    \left[
\begin{array}{cccc}
    0.96 &  \ 0.02&  \ 0.02 & \\   
    0.7125 & \ 2.75 \, cg_{own}^{(1)} + 0.02  &  \ 2.75 (1-cg_{own}^{(1)}) + 0.02 \\ 
    0.7125 &  \ 2.75 (1-cg_{own}^{(2)}) + 0.02  &  \ 2.75 \, cg_{own}^{(2)} + 0.02 \\
\end{array}
\right]
\end{equation}
Let us note here that based on the above matrix of very specific reproduction numbers (from which group to which), we can accumulate the results to arrive to the total number of generated infections for each group. By summing up each line, we see that the expected number of infections caused by an infected GP individual is 1, while that for a CG individual is 3.5. The basic reproduction number for the whole population can be obtained as the spectral radius of this matrix, for example whenever $cg_{own}^{(1)}=cg_{own}^{(2)}$, the spectral radius of this given matrix is approximately $2.8$.

\section{Aims and results}

\subsection{Goals}
In this subsection, we define what goals we have in mind with our simulation study and what sort of consequences we would like to draw from them. As seen from the reproduction numbers \eqref{R0s}, in our parameter setup, just within the general population in the absence of core groups, the disease would  die out (since the reproduction number would become $0.96<1$). Meanwhile, in the core groups it is expected to spread thoroughly. In our simulation, we start the disease from the first core group, where 10 individuals are initially infected. Then spreading takes its course as described in Section \ref{sec:dynamics}.  

We are interested in the effect of restrictions that weaken the bond between the two core groups. In order to see that, we decrease the proportion of sexual connections from one group to the other. However, we take two different approaches to do so, depending on the response to the restriction that disables a certain connection.
\begin{enumerate}
\item Active response or rewiring: the individual whose connection has been disabled redirects it towards someone else. In our case, instead of having a connection from the other core group, it is rewired within its own instead.
\item Inactive response or deletion: the connection is just deleted, and no new connection replaces it.
\end{enumerate}

Here we adapt the terminology active/inactive as it has been introduced before in the context of network epidemics \cite{active}.

\subsection{Results}

In this subsection, we present the results of a simulation study with 3 different base setups and 4 scenarios applied to all of them. First of all, we describe the base setup that we start with in each type of scenario. 

\begin{itemize}
\item \textbf{Base setups:} 
\begin{enumerate}
    \item Joined core groups: since the groups are of equal size, this is equivalent to having $cg_{own}$ parameters of .5.
    \item Symmetric core groups: each group favors itself in terms of sexual connections, with $cg_{own}^{(1)} = cg_{own}^{(2)}  = .8$.
    \item  Asymmetric core groups: each group favors the source (first) core group in terms of sexual connections, with $cg_{own}^{(1)} = 1 - cg_{own}^{(2)}  = 0.8$.
\end{enumerate}
\item \textbf{Scenarios:} we apply restrictions in the form of decreasing the connection between the core groups with $0, 20, 40, 60, 80,$ and $100\%$. The scenarios differ in the effect of the restrictions:
\begin{enumerate}
\item Both Active: Core Group 1 and Core Group 2 both react actively to the restriction, i.e., they form new connections within their own group instead. So, for example, if the proportion of sexual connections to the other group decreases from 0.5 to 0.4, then the proportion of sexual connections to its own group increases from 0.5 to 0.6. More precisely, we modify \eqref{betas} such that we end up with
\begin{align*}
    & \beta_{\mathrm{cg^{1}}}^{\mathrm{cg^{1}}} = c_\textrm{cg}^\textrm{s} p^\textrm{s} ( r + cg_{own}^{(1)} (1 - r) )
        + c^{\textrm{ns}} p^{\textrm{ns}} q_{\textrm{cg}^{1}},
    \\
    & \beta_{\mathrm{cg^{1}}}^{\mathrm{cg^{2}}} = c_\textrm{cg}^\textrm{s} p^\textrm{s} (1-cg_{own}^{(1)})(1-r)
        + c^{\textrm{ns}} p^{\textrm{ns}} q_{\textrm{cg}^{2}},
    \\
    & \beta_{\mathrm{cg^{2}}}^{\mathrm{cg^{1}}} = c_\textrm{cg}^\textrm{s} p^\textrm{s} (1-cg_{own}^{(2)})(1-r)
        + c^{\textrm{ns}} p^{\textrm{ns}} q_{\textrm{cg}^{1}},
    \\
    & \beta_{\mathrm{cg^{2}}}^{\mathrm{cg^{2}}} = c_\textrm{cg}^\textrm{s} p^\textrm{s} ( r + cg_{own}^{(2)} (1 - r) )
        + c^{\textrm{ns}} p^{\textrm{ns}} q_{\textrm{cg}^{2}},
\end{align*}
where $r$ denotes the level of the restriction.
In fact, the total number of sexual contacts for an individual in the core group remains the same for any level of restriction, namely
\[
    c_\textrm{cg}^\textrm{s} ( r + cg_{own}^{(i)} (1 - r) ) + c_\textrm{cg}^\textrm{s} (1-cg_{own}^{(i)})(1-r) = c_\textrm{cg}^\textrm{s}, \qquad i = 1, 2.
\]
\item Both Inactive: Core Group 1 and Core Group 2 both react inactively to the restriction, i.e., no new connections appear in place of the lost ones. So, for example, if the proportion of sexual connections to the other group decreases from 0.5 to 0.4, then the proportion of sexual connections to its own group remains unaffected, 0.5.
More precisely, we change \eqref{betas} to
\begin{align*}
    & \beta_{\mathrm{cg^{1}}}^{\mathrm{cg^{2}}} = c_\textrm{cg}^\textrm{s} p^\textrm{s} (1-cg_{own}^{(1)})(1-r)
        + c^{\textrm{ns}} p^{\textrm{ns}} q_{\textrm{cg}^{2}}, \\
     & \beta_{\mathrm{cg^{2}}}^{\mathrm{cg^{1}}} = c_\textrm{cg}^\textrm{s} p^\textrm{s} (1-cg_{own}^{(2)})(1-r)
        + c^{\textrm{ns}} p^{\textrm{ns}} q_{\textrm{cg}^{1}}.
\end{align*}
\item Active/Inactive: Core Group 1 reacts actively, while Core Group 2 reacts inactively. See previous explanations.
\item  Inactive/Active: Core Group 1 reacts inactively, while Core Group 2 reacts actively. See previous explanations.
\end{enumerate}
\end{itemize}

In the next sections, for each base setup, we provide three sets of plots related to the three groups: the General Group, Core Group 1, Core Group 2. Each plot shows the smoothed empirical distribution of the number of individuals reached by the disease, namely the final size distribution. We performed 1000 simulations in every setup. The smoothing method is a uniform two-sided moving average with window size 21 (except for Figure \ref{fig:distribution_cg2_B3}, where a window size of length 5 was used). Each plot has 6 graphs corresponding to the restriction levels. 

\subsubsection{Base Setup 1.} 
\begin{figure}
    \centering
    \includegraphics[width=\textwidth]{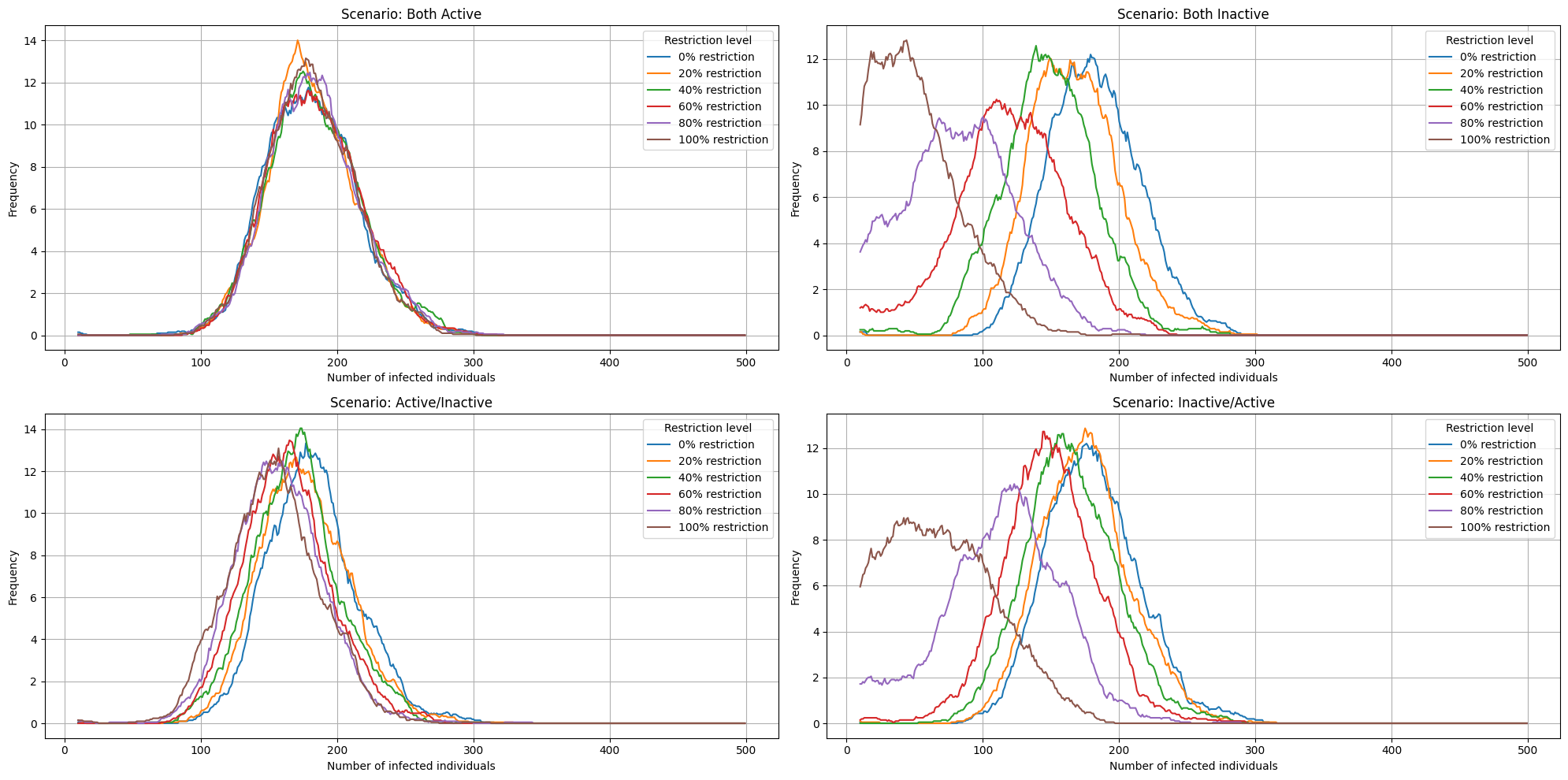}
    \caption{Final size distribution for \textbf{General Group} in \textbf{Base Setup 1} when restrictions are applied on the core groups' sexual connections. The intervention is mostly ineffective in the \textit{Both Active} and \textit{Active/Inactive} scenarios, but has a remarkable influence in the \textit{Both Inactive} and \textit{Inactive/Active} scenarios.}
    \label{fig:distribution_general_B1}
\end{figure}

\begin{figure}
    \centering
    \includegraphics[width=\textwidth]{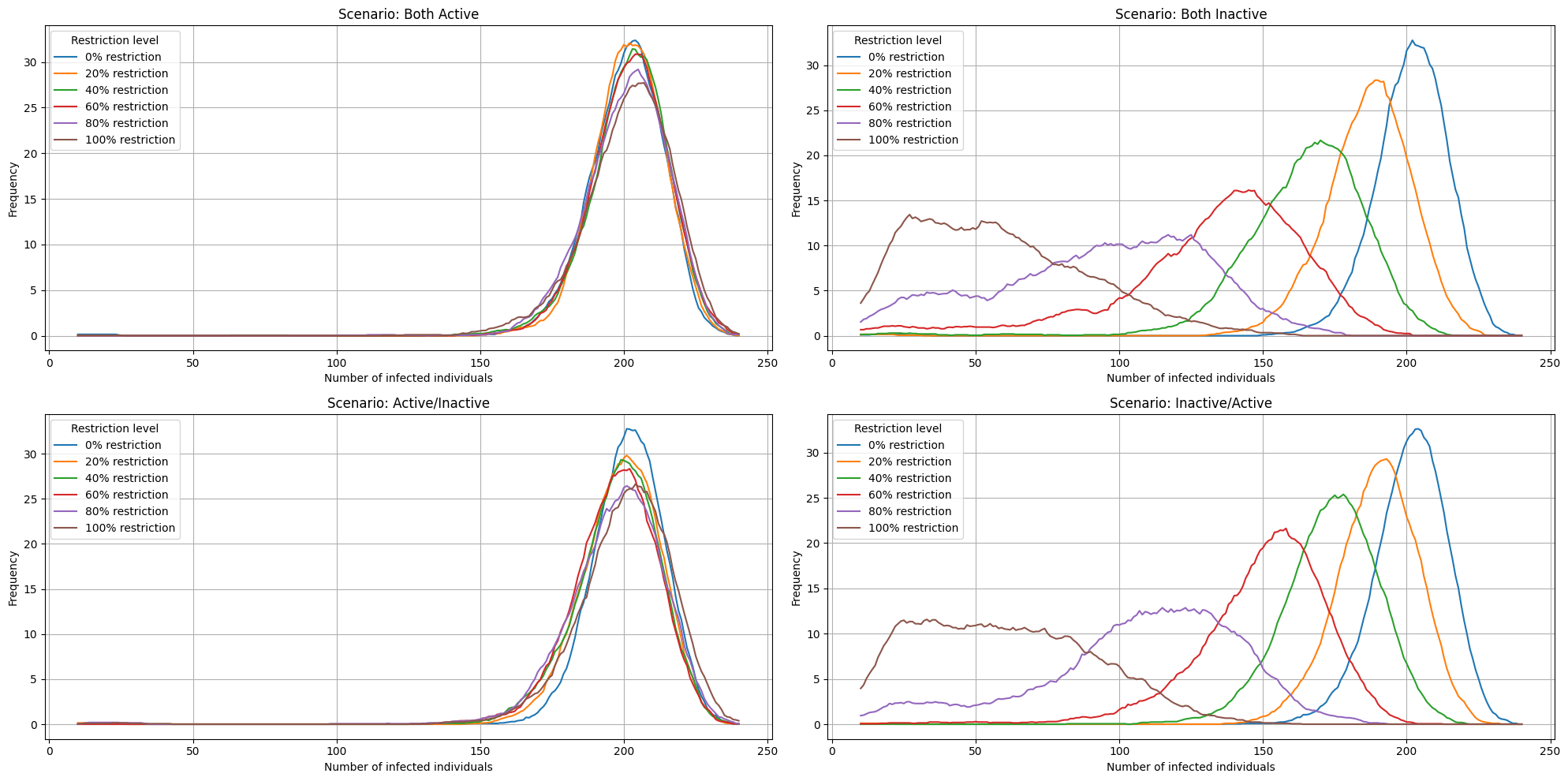}
    \caption{Final size distribution for \textbf{Core Group 1} in \textbf{Base Setup 1} when restrictions are applied on the core groups' sexual connections. The intervention is mostly ineffective in the \textit{Both Active} and \textit{Active/Inactive} scenarios, but has a remarkable influence in the \textit{Both Inactive} and \textit{Inactive/Active} scenarios.}
    \label{fig:distribution_cg1_B1}
\end{figure}

\begin{figure}
    \centering
    \includegraphics[width=\textwidth]{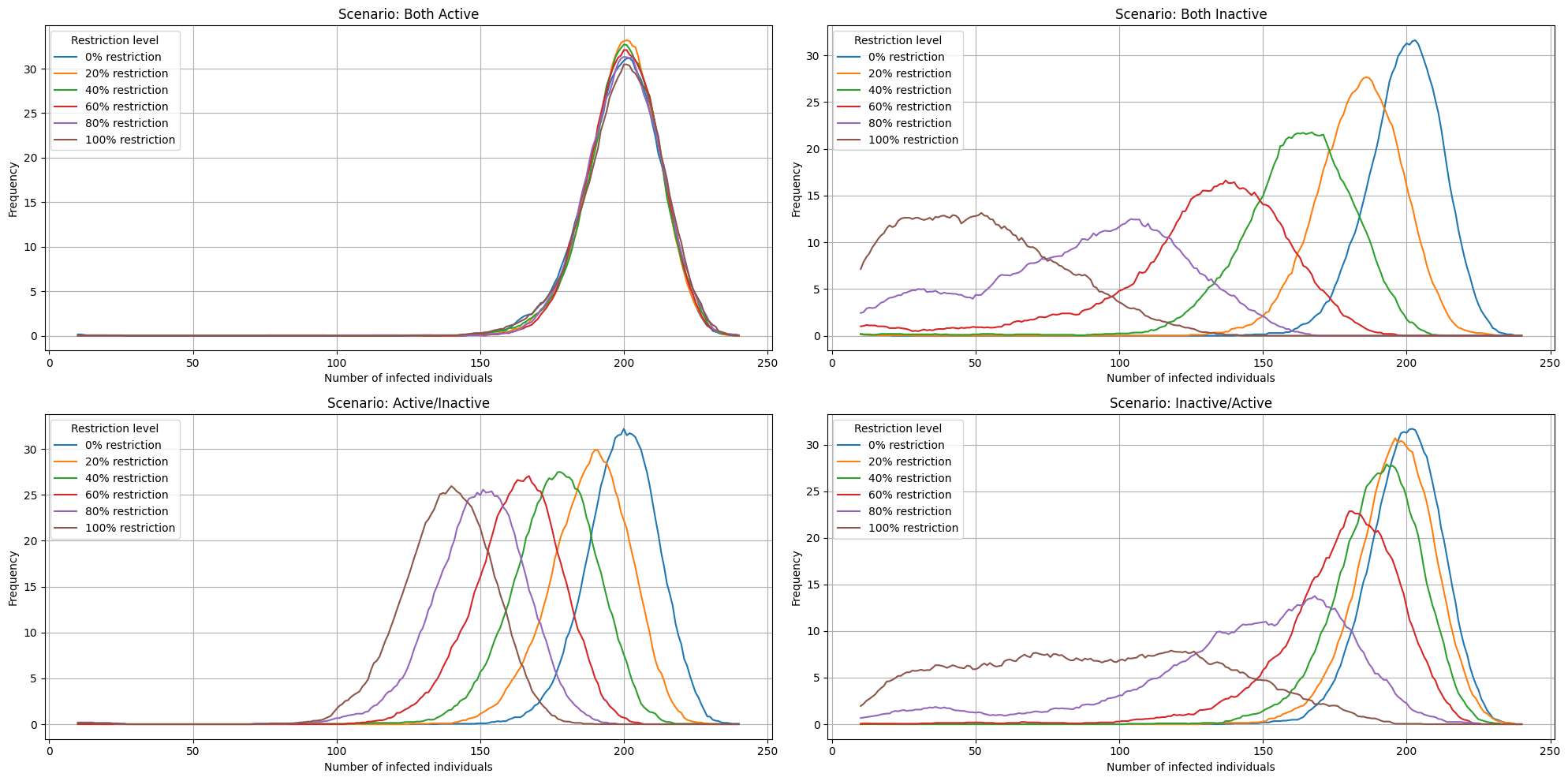}
    \caption{Final size distribution for \textbf{Core Group 2} in \textbf{Base Setup 1} when restrictions are applied on the core groups' sexual connections. The intervention is ineffective in the \textit{Both Active}, moderately effective in the \textit{Active/Inactive} scenario, but has a remarkable influence in the \textit{Both Inactive} and \textit{Inactive/Active} scenarios.}
    \label{fig:distribution_cg2_B1}
\end{figure}

In Figures \ref{fig:distribution_general_B1}, \ref{fig:distribution_cg1_B1}, and \ref{fig:distribution_cg2_B1} 
we can see the results related to the Base Setup 1. In this setup we start with the joined core group case, namely if all the $cg_{own}$ parameters are 0.5, which corresponds to the 0\% restriction case (blue curves). Indeed, the final size distributions are the same in each scenario.
The other curves represent the distributions with the applied restrictions of 20, 40, 60, 80, and 100\%. In these cases, the role of the two core groups is different because, with the intervention, we modify the bond between the core groups. 

For the General Group (Figure \ref{fig:distribution_general_B1}) the intervention is ineffective in the \textit{Both Active} and \textit{Active/Inactive} scenarios, but has a remarkable influence in the \textit{Both Inactive} and \textit{Inactive/Active} scenarios.
A similar situation can be observed for Core Group 1 (Figure \ref{fig:distribution_cg1_B1}).
If the intervention is inactive in a given core group, the number of possible overall infections decreases as the level of restriction increases because connections are deleted. On the other hand, if the intervention is active, then the number of total possible infections remains the same with any level of restriction because the connections are rewired.
The disease starts from Core Group 1, which means that the active or inactive status of that group is crucial, which is what we observed in both Figure \ref{fig:distribution_general_B1} and Figure \ref{fig:distribution_cg1_B1}.
For Core Group 2 (Figure \ref{fig:distribution_cg2_B1}) the effect of interventions similar except in the \textit{Active/Inactive} scenario. There we can see a moderate influence compared to the other groups. The reason is that Core Group 2 is inactive and the disease starts from the other core group.

\subsubsection{Base Setup 2.} 
\begin{figure}
    \centering
    \includegraphics[width=\textwidth]{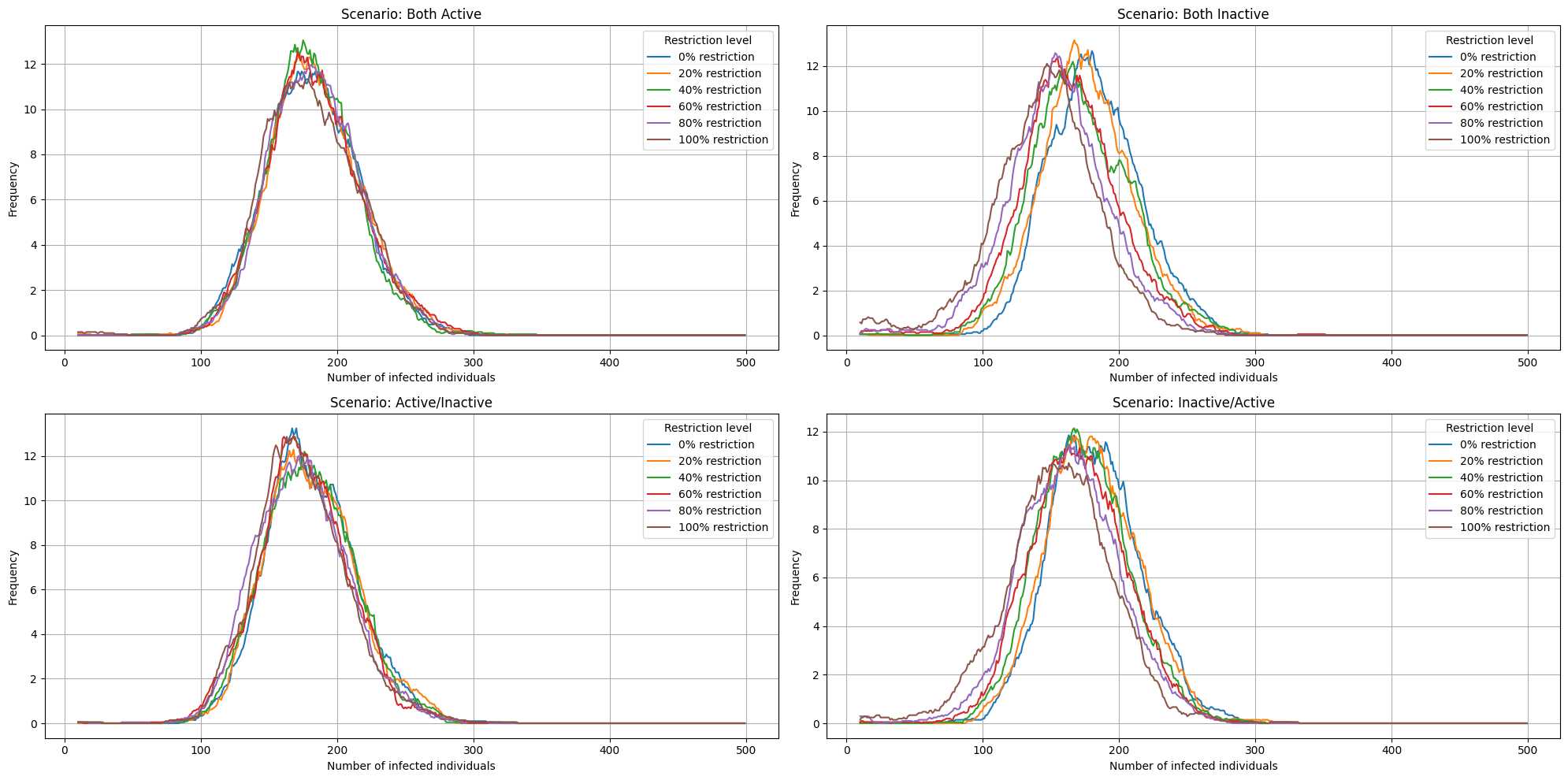}
    \caption{Final size distribution for \textbf{General Group} in \textbf{Base Setup 2} when restrictions are applied on the core groups' sexual connections. The impact of the intervention is negligible.}
    \label{fig:distribution_general_B2}
\end{figure}

\begin{figure}
    \centering
    \includegraphics[width=\textwidth]{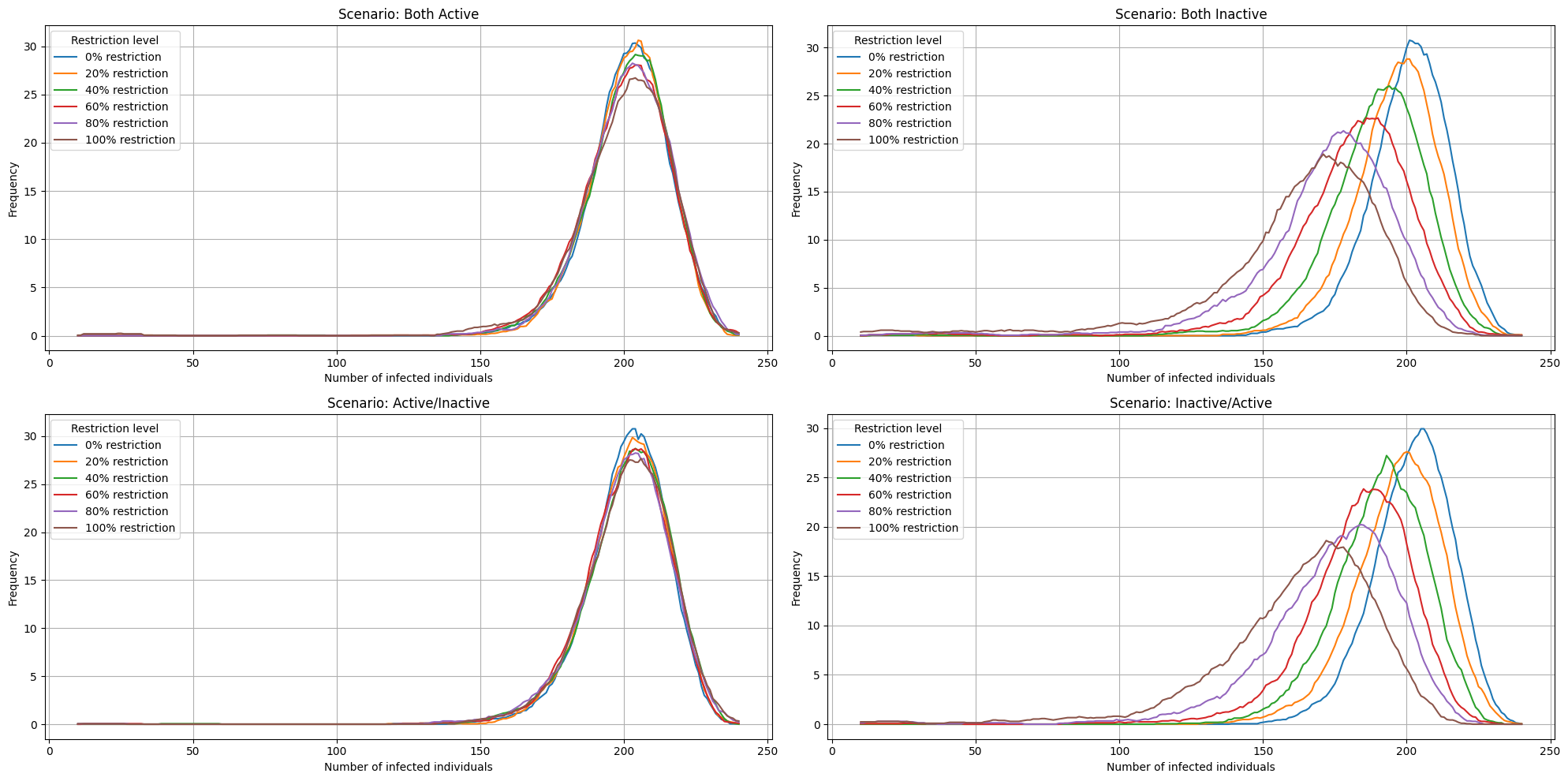}
    \caption{Final size distribution for \textbf{Core Croup 1} in \textbf{Base Setup 2} when restrictions are applied on the core groups' sexual connections. The intervention is ineffective in the \textit{Both Active} and \textit{Active/Inactive} scenarios, but has a moderate effect in the \textit{Both Inactive} and \textit{Inactive/Active} scenarios.}
    \label{fig:distribution_cg1_B2}
\end{figure}

\begin{figure}
    \centering
    \includegraphics[width=\textwidth]{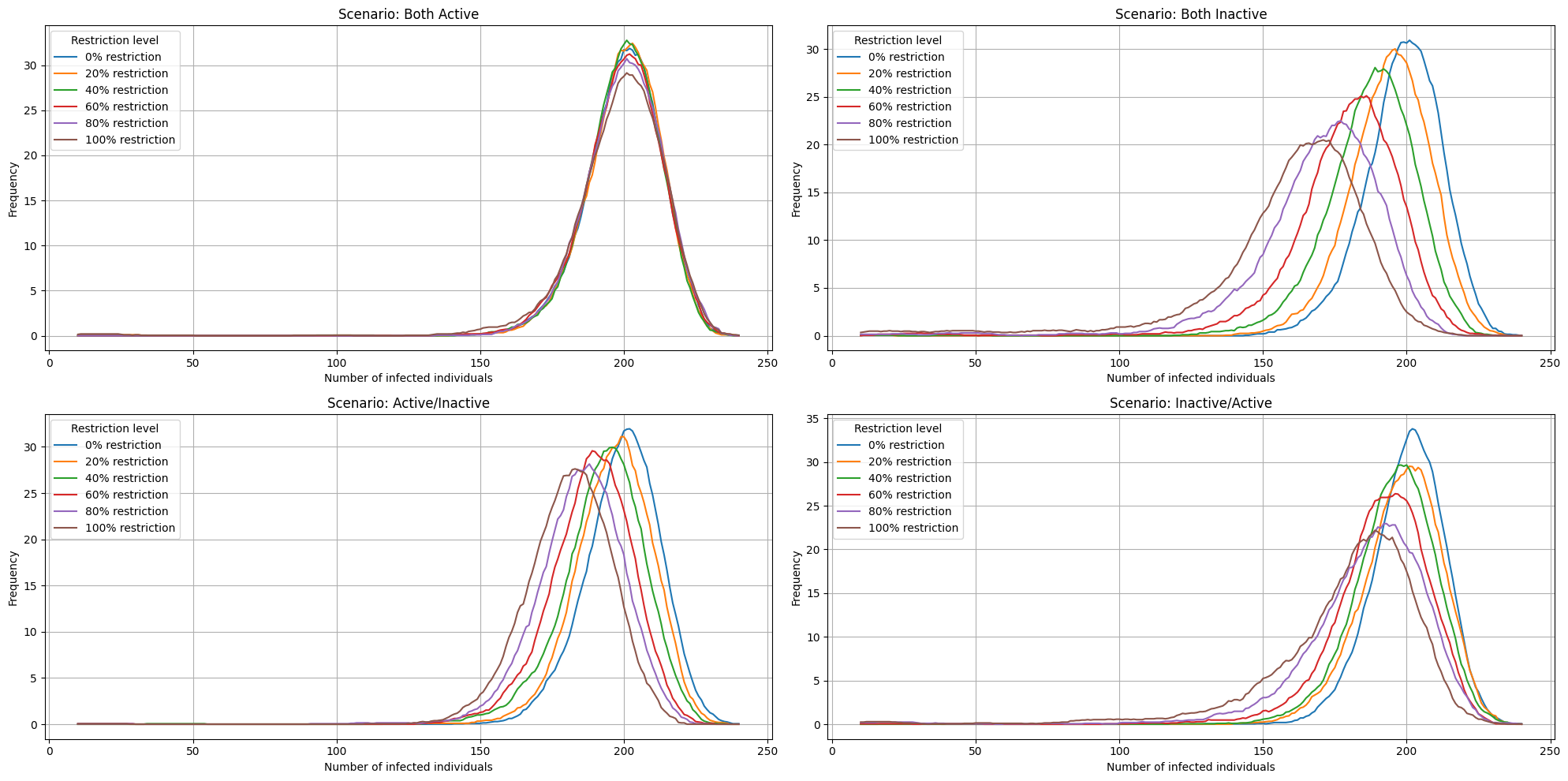}
    \caption{Final size distribution for \textbf{Core Group 2} in \textbf{Base Setup 2} when restrictions are applied on the core groups' sexual connections. The intervention is ineffective only in the \textit{Both Active} scenario and has a partial impact in the remaining scenarios.}
    \label{fig:distribution_cg2_B2}
\end{figure}

In Figures \ref{fig:distribution_general_B2},\ref{fig:distribution_cg1_B2}, and \ref{fig:distribution_cg2_B2} we can see the results related to the Base Setup 2. This setup corresponds to the case of symmetric core groups with $cg_{own}^{(1)} = cg_{own}^{(2)} = 0.8$, namely 80\% of the connection of a core individual points to the own core group and 20\% to the other core group. With the intervention, we weaken the bond between the core groups either in an active or inactive way. The blue curves represent the baseline case (0\% restriction), thus the corresponding distributions are the same in each scenario. 

For the General Group (Figure \ref{fig:distribution_general_B2}), we can see that the impact of the intervention is negligible, with a minimal effect in the \textit{Both Inactive} and in the \textit{Inactive/Active} scenarios. The connection between the core groups was 20\% symmetrically, which we weakened during the interventions, so the infections to the own group are more dominant compared to the cross infections for all levels of restrictions. Therefore, the final size in the General Group remains the same.

In case of the core groups, we can observe some influence of the interventions. For Core Group 1 (Figure \ref{fig:distribution_cg1_B2}), the intervention is ineffective in the \textit{Both Active} and \textit{Active/Inactive} scenarios, but has a moderate effect in the \textit{Both Inactive} and \textit{Inactive/Active} scenarios. In the latter cases Core Group 1 is inactive, meaning that some connection are deleted, which causes a decrease in the final size. 
For Core Group 2 (Figure \ref{fig:distribution_cg2_B2}), we can observe a similar situation. The intervention is only ineffective in the \textit{Both Active} scenario and has a partial impact in the remaining scenarios.

\subsubsection{Base Setup 3.} 
\begin{figure}
    \centering
    \includegraphics[width=\textwidth]{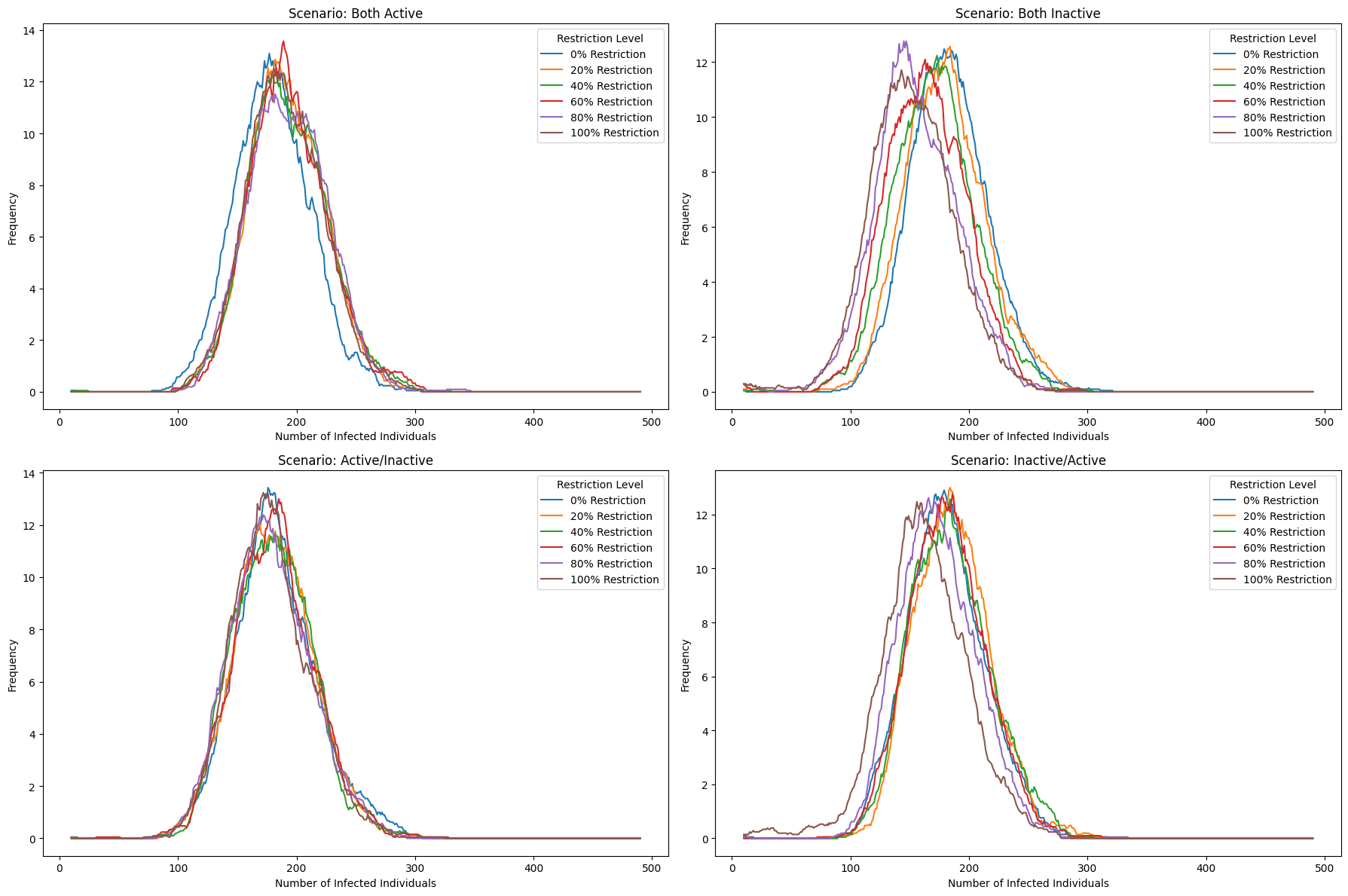}
    \caption{Final size distribution for \textbf{General Group} in \textbf{Base Setup 3} when restrictions are applied on the core groups' sexual connections.
    In the \textit{Both Active} and in the \textit{Active/Inactive} scenarios the intervention is ineffective, while in the \textit{Both Inactive} and in the \textit{Inactive/Active} scenarios it has a moderate impact on the final size.
    }
    \label{fig:distribution_general_B3}
\end{figure}

\begin{figure}
    \centering
    \includegraphics[width=\textwidth]{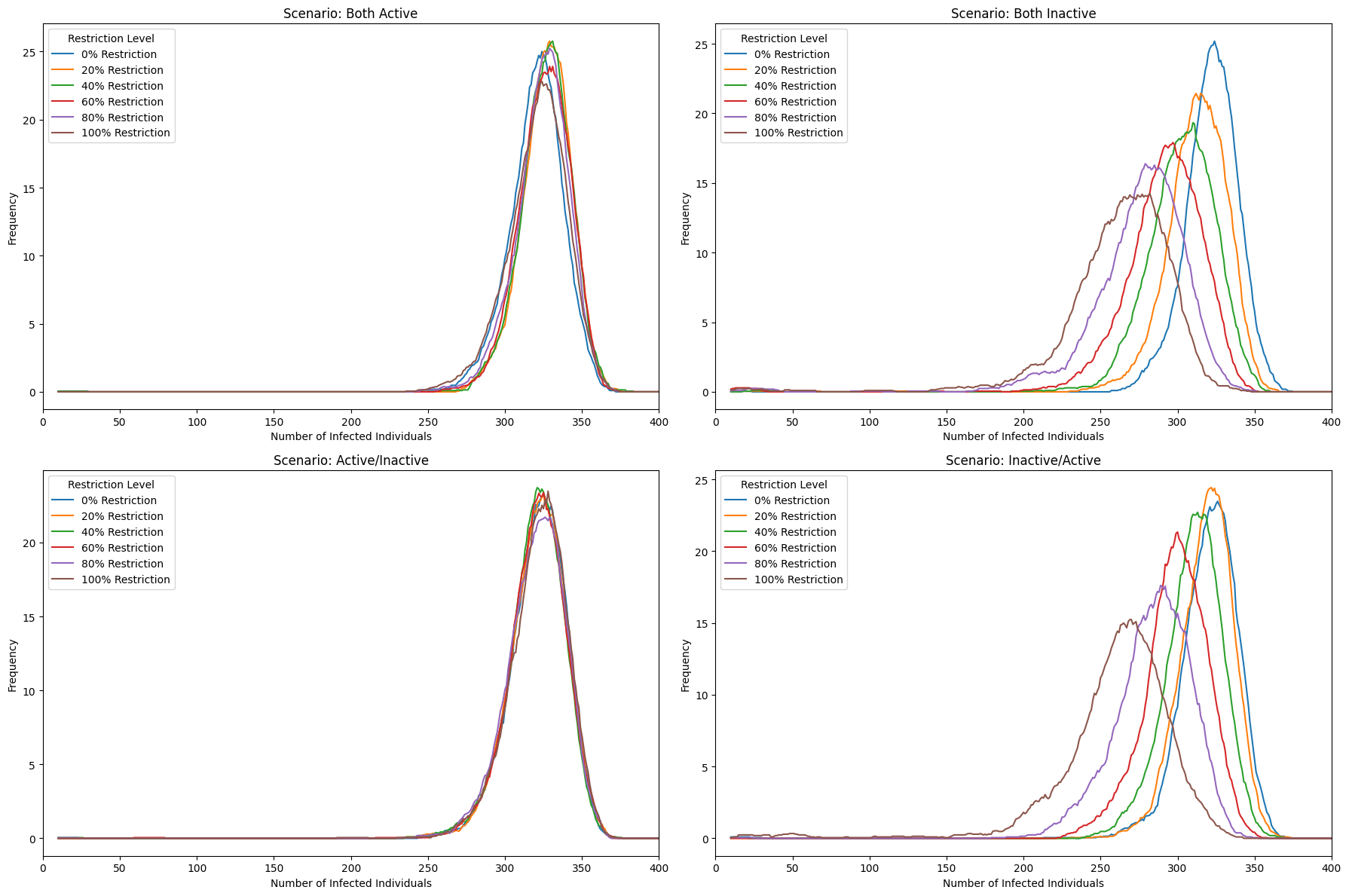}
    \caption{Final size distribution for \textbf{Core Group 1} in \textbf{Base Setup 3} when restrictions are applied on the core groups' sexual connections. The final size has been reduced well in the \textit{Both Inactive} and in the \textit{Inactive/Active} scenarios, and we can see stagnation in \textit{Both Active} and \textit{Active/Inactive} scenarios.}
    \label{fig:distribution_cg1_B3}
\end{figure}

\begin{figure}
    \centering
    \includegraphics[width=\textwidth]{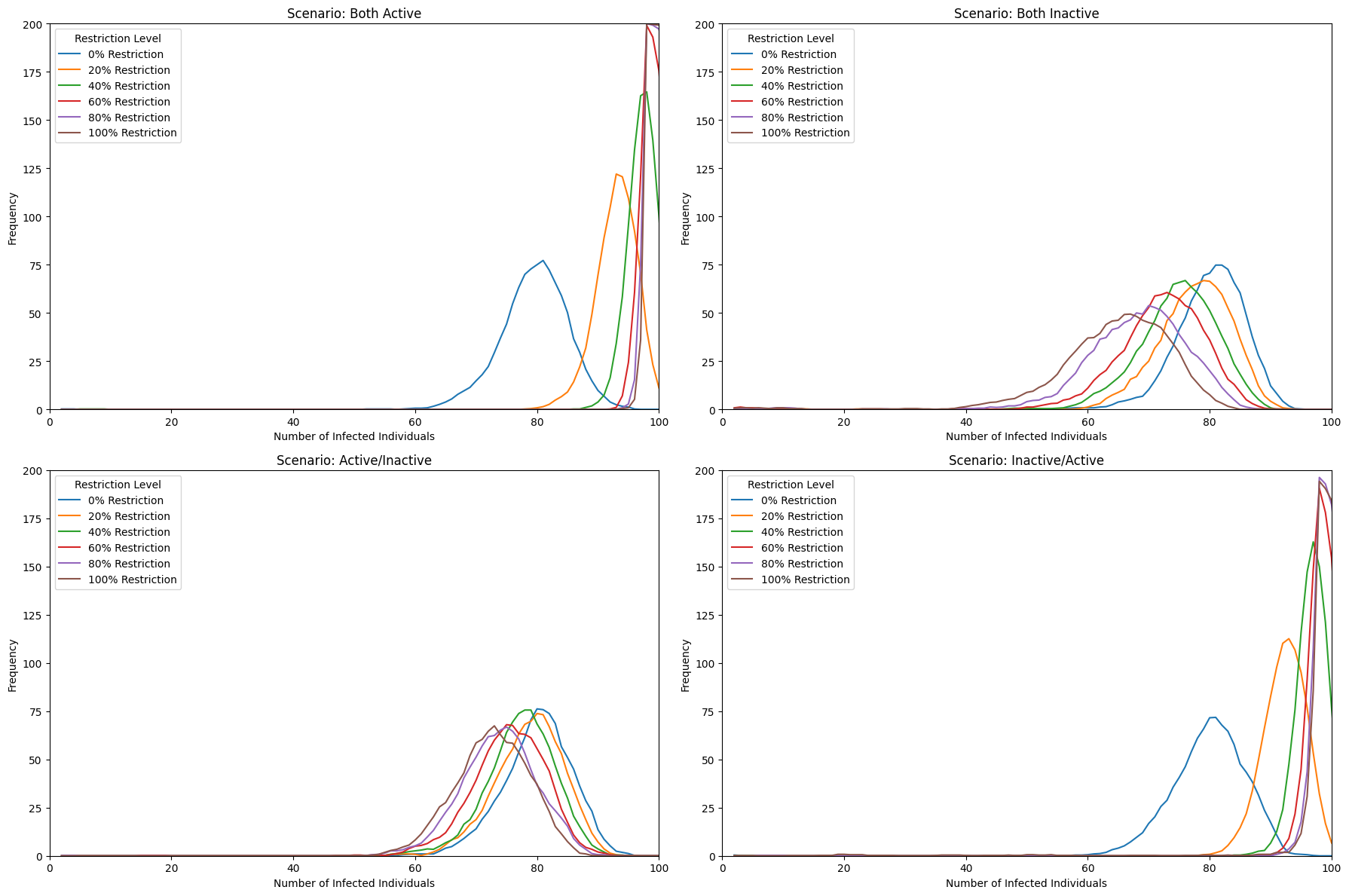}
    \caption{Final size distribution for \textbf{Core Group 2} in \textbf{Base Setup 3} when restrictions are applied on the core groups' sexual connections. In the \textit{Both Inactive} and \textit{Active/Inactive} scenarios the final size is effectively decreased, but in the \textit{Both Active} and \textit{Inactive/Active} scenarios the final size is considerably increased.}
    \label{fig:distribution_cg2_B3}
\end{figure}

In Figures \ref{fig:distribution_general_B3}, \ref{fig:distribution_cg1_B3}, and \ref{fig:distribution_cg2_B3} we can see the results related to the Base Setup 3. This one corresponds to the case of asymmetric core groups with $cg_{own}^{(1)} = 1 - cg_{own}^{(2)} = 0.8$, namely Core Group 1 has a dominant role compared to Core Group 2, because 80\% of the connection of a core individual points to Core Group 1, independently of the origin of that core individual.
Moreover, in this setup we consider different core group sizes, see the explanation in Section \ref{sec:par}.
With the intervention, we weaken the bond between the core groups either in an active or an inactive way. The blue curves represent the baseline case (0\% restriction), thus the corresponding distributions are the same in each scenario.

For the General Group (Figure \ref{fig:distribution_general_B3}), we can observe that the intervention has a mild effect on the final size. In the \textit{Both Active} and in the \textit{Active/Inactive} scenarios the intervention is ineffective, while in the \textit{Both Inactive} and in the \textit{Inactive/Active} scenarios it has a moderate impact on the final size. 

In case of the two core groups, we can see different influence of the interventions.
For Core Group 1 (Figure \ref{fig:distribution_cg1_B3}) the final size has been reduced well in the \textit{Both Inactive} and in the \textit{Inactive/Active} scenarios, and we can see stagnation in \textit{Both Active} and \textit{Active/Inactive} scenarios.
For Core Group 2 (Figure \ref{fig:distribution_cg2_B3}), we can see an unexpected situation.
In the \textit{Both Inactive} and \textit{Active/Inactive} scenarios the final size is effectively decreased, but in the \textit{Both Active} and \textit{Inactive/Active} scenarios the final size is considerably increased.
The reason is the following. If Core Group 2 has an active role, then during the intervention Core Group 1 loses its dominance, because the cross connection is decreased between the core groups. Therefore, due to the active response, the connection to its own group is increased, implying a greater amount of final size in Core Group 2. 

\section{Discussion}
In this study, a stochastic SEIR model is introduced for a structured population with a general and two core groups. The main aim is to investigate the effect of commonly applied non-pharmaceutical interventions in which we weaken the bond between core groups. We have considered three base setups with four different scenarios. Based on the results, we can deduce that the connection structure between the core groups is crucial to choose the most effective intervention.

In Base Setup 1, in which the core groups have similar characteristics, we find that an intervention can be efficient only if at least one of the core groups reacts inactively (without forming new connections to replace of the lost ones) to the intervention. Therefore, we have to avoid the rewiring of the connections to make the intervention effective.
Base Setup 2 is similar to the first one, but here each core group favors itself in terms of sexual connections over the other core group (assortative mixing). The conclusion is analogous, to maximize the efficiency of the intervention we have to prevent the rewiring response. Furthermore, it is worth mentioning that, due to the symmetry of the core groups, these interventions are ineffective for the General Group.
Finally, in Base Setup 3 we consider the case when the core groups have different roles in the connection structure. In fact, the same proportion of a core individual's sexual connections are directed towards Core Group 1, regardless of the origin of the individual. It turned out that an intervention is efficient if the less influential core group is inactive. However, if this secondary core group has an active response, then the intervention leads to a moderately worse outcome than no intervention.
Therefore, if the core groups have asymmetric connection structure, we have to pay more attention on the rewiring response to the intervention. 

\section*{Acknowledgement}
JB was supported by the National Research, Development and Innovation Office (NKFIH) in Hungary, grant RRF-2.3.1-21-2022-00006 and TKP2021-NVA-09.  NF was supported by TKP-2021-EGA-05, EU H2020 No. 739593, and 2022-2.1.1-NL-2022-00005. GR was supported by KKP 129877 and 2024-1.2.3-HU-RIZONT-2024-00034.

\bibliographystyle{unsrt}  
\bibliography{references}   

\end{document}